\documentclass{article}
\usepackage[preprint]{neurips_2019}

\usepackage[utf8]{inputenc} 
\usepackage[T1]{fontenc}    
\usepackage{hyperref}       
\usepackage{url}            
\usepackage{booktabs}       
\usepackage{amsfonts}       
\usepackage{nicefrac}       
\usepackage{microtype}      
\usepackage{color, colortbl} 

\title{Comparing manual contact tracing and digital contact advice}

\author{%
  Ramesh Raskar\\
  MIT Media Lab/Pathcheck
  \And
  Dr. Ranu Dhillon\\
  Harvard Medical School
  \And
  Dr. Suraj Kapa\\
  Mayo Clinic
  \And
  Deepti Pahwa\\
  Stanford GSB
  \And
  Renaud Falgas\\
  The Swatch Group
  \And
  Lagnojita Sinha\\
  MIT Media Lab
  \And
  Aarathi Prasad\\
  Skidmore College
  \And
  Abhishek Singh\\
  MIT Media Lab
  \And
  Andrea Nuzzo\\
  GSK
  \And
  Rohan Iyer\\
  Pathcheck
  \And
  Vivek Sharma\\
  MIT Media Lab \\Harvard Medical School
}

\begin{document}

\maketitle

\begin{abstract}
  Manual contact tracing is a top-down solution that starts with contact tracers at the public health level, who identify the contacts of infected individuals, interview them to get additional context about the exposure, and also monitor their symptoms and support them until the incubation period is past. On the other hand, digital contact tracing is a bottom-up solution that starts with citizens who on obtaining a notification about possible exposure to an infected individual may choose to ignore the notification, get tested to determine if they were actually exposed or self-isolate and monitor their symptoms over the next two weeks. Most experts recommend a combination of manual contact tracing and digital contact advice but they are not based on a scientific basis. For example, a possible hybrid solution could involve a smartphone based alert that requests the possible contact of an infected individual to call the Public Health (PH) number for next steps, or in some cases, suggest ways to self-assess in order to reduce the burden on PH so only most critical cases require a phone conversation.  In this paper, we aim to compare the manual and digital approaches to contact tracing and provide suggestions for potential hybrid solutions.
\end{abstract}

\section{Introduction}

Manual contact tracing (MCT) involves building an ‘army of detectives’ that perform three steps as per WHO (i) contact identification: quickly backtraces the contacts of an infected individual, (ii) interviews them and (iii) follow up on changes in their symptoms. During the interview, the contact tracers may also identify potential surface or regional exposures to identify other routes of transmission (eg, shared surfaces, shared equipment, etc). Contact tracers have historically been critical to fighting a pandemic as contact tracing can be crucial when determining how a new pathogen, such as COVID spreads. At Safe Paths we aim to frame digital capabilities in the context of an overall ‘ideal’ contact tracing system. We believe there ultimately may be advantages/functionalities of digital tools to enhance certain aspects of Manual Tracing. For example, according to CDC recommendations, case investigators may ask infected individuals to refer to location-tracking apps to determine their recent location history. 

Digital contact advice (DCA) is a relatively new field. DCA software may perform one or more of the following tasks: 
\begin{itemize}
  \item Identify contacts of infected individuals
  \item Notify them when their exposure time window expires
  \item Provide an initial survey about their symptoms and clear instructions on how to regularly monitor their symptoms and health status and report that information every day
  \item Provide public safety messages to contacts to educate them about COVID-19
\end{itemize}
Smartphone apps that use location logging (e.g. GPS) or proximity logging (e.g. bluetooth) technologies to alert a smartphone user whether they were in contact with an infected individual retrospectively have also become popular.  The effectiveness of these digital technologies for contact tracing are still being validated. Additionally, manual tracing's historic efficacy is often used to challenge digital contact tracing without acknowledging some of the risks that are associated with the manual approach. \cite{Samarajiva} \cite{Criteria}

Public safety messages to identified contacts to educate them about COVID-19, its common signs and symptoms, and reinforcing prevention messages defined by the government, such as self-quarantine and social distancing. (This messaging should be repeated daily throughout the contact’s self-quarantine period with new information supportive of the evolving stage of isolation.) 

Ability to send notifications in multiple formats, such as voice messages, emails, and SMS.

Capability for contact-generated and system-generated alerts or workflows (e.g., to facilitate appropriate follow-up, presence of symptoms, contact request for information). \cite{Salesforce}

Ability to produce individual-level and aggregate data supporting worker and PHA-level process metrics as described above. 

In the following sections, we present some key differences in MCT and DCA.

\section{Metrics}

\textbf{Speed}: Time taken to find contacts through manual tracing is variable. If a patient is sick during the interview, it might take longer for the contact tracer to identify close contacts because the patient may not be capable of recollecting their past encounters. It may also require multiple interviews to jog the memory of the infected individual.

With digital tracing, a contact could obtain an exposure notification as soon as an infected individual discloses their location or proximity logs to other users or send their diagnosis to a centralized server. This instantaneous notification would let people know they may have been exposed and immediately isolate to preclude pre-symptomatic/asymptomatic spread.

\textbf{Integrity}: Integrity is a very important factor in the whole contact tracing system given that the success hinges on it to a significant extent. One way of interpreting integrity of the contact tracing pipeline is the extent of false positives and false negatives. In Guinea, (during EBOLA outbreak) HAs used to have people not give names of people who were wage-earners to avoid them not being able to work. Some who didn’t trust the response would not give a real list of contacts. MCT can miss many contacts. People, especially when ill, won’t remember everywhere they’ve been and also would never know the names or numbers of strangers who they may have exposed (eg, on a subway). So the false negative is higher in the manual contact tracing system because many exposed individuals get missed out. On the other side, digital contact tracing (DCT) the rate of false positives are the biggest issue - particularly people who were within a certain distance but behind a wall or neighboring apartment or any physically separated boundary which inhibits the transmission of airborne particles. Some of the false positives and negatives in DCT could also arise from the general distrust of systems like this.

\textbf{Effectiveness}: MCT has been shown to be very effective with past pandemics such as SARS, HIV, Ebola and Zika. A trained tracer’s ability to assess symptoms, detect asymptomatic carriers, and flag other possibilities for compromised health and onward spread, which untrained eyes and apps may struggle to identify. \cite{Bailey} \cite{WHO}

However, effectiveness depends highly on the region, and the community. Also, only about 50\% of contacts may pick up the phone when tracers try to reach out to them and it is unclear how many follow the procedures suggested by the tracers.

With DCA, effectiveness depends significantly on adoption. The smartphone apps may only be downloaded by tech-savvy users, and not by vulnerable populations such as elderly, and low-income communities who cannot afford the smartphones that have the technologies required for location and proximity detection (cite?). Concerns of privacy also prevent people from downloading location tracking apps. All software involved should be interoperable, to receive input from the public health authorities (PHA) (including local, state, tribal, and territorial public health departments), information systems and/or laboratory systems, either via import or real-time synchronization.

Fundamentally, contact tracing works by tracking down all the contacts of an infected person and then taking appropriate action to break the chain of transmission.  Case management tools for case investigation and contact tracing capture data on cases and contacts and can help improve the efficiency of manual contact tracing and medical monitoring methodologies. In order for digital contact tracing to be effective and a good supplement to manual contact tracing, it should generally have the following capabilities \cite{Tools}:
\begin{itemize}
  \item Ability to ensure data security and confidentiality of significant volumes of client information, which is critical to maintain community trust in using any case management tool.
  \item Interoperability capabilities to receive input from the public health authorities (PHA) (including local, state, tribal, and territorial public health departments), information systems and/or laboratory systems, either via import or real-time synchronization.
  \item Ability to facilitate identification/elicitation and documentation of known contacts of clients with COVID-19, both through manual entry by the PHA and via self-report from cases.
  \item Ability to send notifications to users (clients and contacts) via manual and/or automated means. These messages will include:
  \begin{itemize}
    \item Notification to contacts of their exposure and time window when exposure may have occurred.
    \item Initial survey about their symptoms and clear instructions on how to regularly monitor their symptoms and health status and report that information every day. (This will ensure their data reaches the contact management team at the PHA and that aggregate data reach relevant state and federal partners.)
    \item Public safety messages to identified contacts to educate them about COVID-19, its common signs and symptoms, and reinforcing prevention messages defined by the government, such as self-quarantine and social distancing. (This messaging should be repeated daily throughout the contact’s self-quarantine period with new information supportive of the evolving stage of isolation.)
  \end{itemize}
  \item Ability to send notifications in multiple formats, such as voice messages, emails, and SMS.
  \item Capability for contact-generated and system-generated alerts or workflows (e.g., to facilitate appropriate follow-up, presence of symptoms, contact request for information).
  \item Ability to produce individual-level and aggregate data supporting worker and PHA-level process metrics as described above. 
\end{itemize}

\textbf{Scale}:  The scale of operations increase linearly with the increase in the target population in the MCT technology. This can introduce burden during the first phase of the outbreak when the cases increase exponentially. The procedures like training of the staff and carrying out formalities of the job can create a bottleneck during the surge of the cases. In digital contact advice, the scale of the operations do not scale linearly with the target population or users because the same IT infrastructure can serve multiple individuals at the same time due to the technologies like elastic scalable cloud services offering.

\textbf{Cost}: This will ultimately depend on the number of cases and how many contacts need to be followed per day so could become very different if cases surge. There would be a lag time in scaling up/training more tracers. When there is exponential growth in transmission, it’s tough to scale the operations and associated infrastructure commensurately. There are several other factors of cost associated with manual contact tracing independent from technologies and limited network needs (Cost of cell phone credits and printing all forms needed in the field); hand sanitizer; transportation: taxis, fuel (if using own vehicles) or rental vehicles (if needed). However, the DCA does not require proportional growth in infrastructure and hence the cost is more like one-time investment which is required during the setup of the software development and maintenance and the cloud services needed for the DCA process to work.

\textbf{Empathy and benevolence}: Tracers needed who can engender trust to get people to disclose sensitive personal information over the phone, there is no evidence that newly trained tracers are successful. Most manual contact tracers are local entities and this knowledge might influence their social behavior even with those who are related to the exposed. DCTT lacks a range of  human capabilities and characteristics such as to clarify misconceptions, address worries, and express. \cite{Kahn}

\pagebreak

\definecolor{Blue}{rgb}{0.62, 0.77, 0.91}

\begin{table}[!ht]
    \centering
    \begin{tabular}{|p{0.2\textwidth}|p{0.4\textwidth}|p{0.4\textwidth}|}
    \hline
    & Manual contact tracing & Digital contact advice\\
    \hline
    Speed
    \cellcolor{Blue}
    & The speed is variable and highly dependent on the severity of the infected individual and also on the range of number of cases in a given jurisdiction.
    &
    Speed is one of the big advantages of digital contact tracing allowing for a real-time collection and processing of the required data.
    \\
    \hline
    Integrity (false positives and negatives)
    \cellcolor{Blue}
    &
    The false negatives (missed cases) could be higher and it  could vary depending upon other contexts (severity of infected person)
    &
    The false positive rate is higher in the digital contact tracing system.
    \\
    \hline
    Effectiveness
    \cellcolor{Blue}
    &
    About 50\% of contacts pick up the phone and unclear how many follow the procedures; highly dependent on the region, community, etc. MCT has some limited evidence regarding the effectiveness of manual contact tracing, which is lacking for DCA \cite{Kahn}
    &
    Effectiveness rests upon multiple factors here - Accuracy, reliability, and adoption.
    Accuracy - Depends upon the technology used. For bluetooth, the accuracy is sufficient for proximity sensing.
    Reliability - Reliability varies based on the OS of the phone. iOS has some known issues with consistency and background processing.
    Adoption - Highly effective only if most of the population in the community adopt DCA. If only a very small fraction of the community uses DCA it may lead to a false notion of safety.
    \\
    \hline
    Cost
    \cellcolor{Blue}
    &
    Capital: Setting up call centers, Cost of training contact tracers, CRM software
    Operations: Salaries and other costs associated with maintenance.
    Limited financial resources are needed compared to DCA applications. Technicals equipment is already available and tested.
    &
    Capital: Development of software
    Operations: Network and storage (current estimate for US wide server/bandwidth cost is \$50K/day)
    \\
    \hline
    Scale
    \cellcolor{Blue}
    &
    The recommended numbers are 1 Contract Tracer for 30K population. Cost of 100K tracers in the US for one year is \$6B. Maintenance cost scales linearly with the target population.
    &
    Requires a significant part of the population to participate: smartphone usage, app usage, consent to GPS/BT. The IT cost does not scale linearly with the population.
    \\
    \hline
    Empathy and Benevolence
    \cellcolor{Blue}
    &
    Human-to-human phone call provides a rich connection, but depends on training of the tracer. Need to convince infected individuals to reveal details of their activities.
    &
    Lacks empathy. Infected individuals are not in a state of mind to ‘press a button’ instantly to inform all the contacts. In a relatively hybrid approach it can be facilitated by PH officials.
    \\
    \hline
    \end{tabular}
    \label{tab:metrics}
\end{table}
\section{Understanding Context and Triage}

\textbf{2\textsuperscript{nd} degree contacts (contact of contact)}: Manual contact tracers interact with individuals who are confirmed or suspected cases and contacts of cases, but not other members of the general public. It is possible to determine 2\textsuperscript{nd} degree contacts, if they interview the exposed contacts. On the other hand, digital notifications will be sent to even 2\textsuperscript{nd} degree contacts. \cite{Congress}

\textbf{Non-direct methods of transmission}: Shared surfaces may be identified through interviews. On the other hand, for digital approaches, it depends on the wireless technology used for contact tracing - when using GPS or WiFi for location tracking, it may be possible to provide coverage for the non-direct methods but proximity-based techniques may need changes to the infrastructure. 

\textbf{Self-assessment and questionnaires}: For the manual approach, contact tracers will provide questionnaires over the phone, so it is prone to errors. Alternatively, using smartphones could enable higher engagement on self-assessment by the user due to easy access and lower barriers of reporting, and seem less intrusive. \cite{Bode}

\textbf{Triage and next steps}: Triage and providing support to close contacts are possible with both approaches - for the manual approach, it would require effort from the contact tracer, while for digital approach, it is possible to generate personalized questionnaires through the smartphone app. 

\textbf{Follow-up after contact tracing}: For the manual approach, it would require effort from the contact tracer, while for digital approach, it would require the patients to report their information through the app.

\definecolor{Orange}{rgb}{1, 0.85, 0.40}

\begin{table}[!ht]
    \centering
    \begin{tabular}{|p{0.2\textwidth}|p{0.4\textwidth}|p{0.4\textwidth}|}
    \hline
    & Manual contact tracing & Digital contact advice\\
    \hline
    2\textsuperscript{nd} degree contacts (contact of contact)
    \cellcolor{Orange}
    & Possible to determine (but difficult) if contact tracers interview close contacts
    &
    Digital intervention would affect all users regardless of circumstances (though some more than others). 
    \\
    \hline
    Non-direct methods of transmission
    \cellcolor{Orange}
    &
    Shared surfaces may be identified particularly in communal living outbreaks.
    &
    Will not work without changes to wireless infrastructure. It depends on the mechanism used for contact tracing. If GPS is used then it might be possible to provide coverage for the non-direct methods.
    \\
    \hline
    Self-assessment and questionnaires
    \cellcolor{Orange}
    &
    Must be performed over phone with risk of errors
    &
    Use of tech/smartphones can enable higher engagement and seem less intrusive.
    \\
    \hline
    Triage and next steps
    \cellcolor{Orange}
    &
    Performed with confidence, dignity and likelihood of followup if tailored to individual
    &
    Smartphone apps could include personalized follow-up questionnaires. These questions could trigger PH officials to reach out to those with certain needs.
    \\
    \hline
    Follow-up after contact tracing
    \cellcolor{Orange}
    &
    Facilitating the connection of laboratory-confirmed patients with services needed to support a 14-day self-isolation process (e.g., safe housing, food) Regular calling is needed.
    &
    Can Enable patients to report their validated testing status, relevant demographic data, data facilitating the connection with supportive services, and serve as the best means of communication. \\
    \hline
    \end{tabular}
    \label{tab:understanding}
\end{table}
\section{Technology}

\textbf{Need for internet connectivity}: Manual contact tracing does not require any internet connectivity as it is purely based on individuals providing voluntary information: name, address and phone number. While for the digital  approach, one requires reliable internet connectivity, such as for smart-phone apps, technology infrastructure is needed. But if we envision the technology in a way that could work without phones, e.g. QR Codes, Alerts systems, hotspot detections, etc. it could still provide a reasonable amount of benefits.

\textbf{Need for centralized servers}: Manual contact tracing doesn't require any centralized servers, still it requires careful reflections on these areas: How does the data get from the call to some usable medium? Does the tracer take notes and then transfer them into an online survey? What happens to the notes? What happens to the survey? Some places use enterprise class CRM for the results. On the other hand, digital tracing needs to store information collected by smartphones, for processing data for health officials or for merely sending data from infected individuals to other app users, without storing it. For digital approach, servers need not to be centralized, but are the most common and viable solution.

\textbf{Workforce as contact tracers}: Manual contact tracing needed for recruiting and training huge numbers of human contact tracers. On the other end, to be most effective for digital approach, human contact tracers would be needed in the loop only at fewer intersections to conduct follow-up interviews. \cite{SaferMe}

\textbf{Room for error}: Manual contact tracing is based on human memory and thus is prone to potentially significant error in the form of data loss. In contrast, digital contact advice is centered around technology-driven data collection which cannot ‘forget’ and is thus less prone to such data loss. \cite{Gidari}

\definecolor{Gray}{rgb}{0.82, 0.88, 0.89}

\begin{table}[!ht]
    \centering
    \begin{tabular}{|p{0.2\textwidth}|p{0.4\textwidth}|p{0.4\textwidth}|}
    \hline
    & Manual contact tracing & Digital contact advice\\
    \hline
    Need for internet connectivity
    \cellcolor{Gray}
    &
    Not required
    &
    For smart-phone apps, technology infrastructure is needed.  
    \\
    \hline
    Need for centralized servers
    \cellcolor{Gray}
    &
    Not required, but still require human contact tracers in the loop to conduct follow-up.
    &
    Servers need to store information collected by smartphones, with privacy.
    \\
    \hline
    Workforce as contact tracers
    \cellcolor{Gray}
    &
    Need for recruiting and training huge number of contact tracers
    &
    Contact Tracers/ Humans would be needed at fewer intersections. 
    \\
    \hline
    Room for error
    \cellcolor{Gray}
    &
    Manual tracing opens room for errors in recording due to the fallacy of human memory.
    &
    Technology provides an automatic error free solution. 
    \\
    \hline
    \end{tabular}
    \label{tab:technology}
\end{table}
\section{Privacy}

In this section, we consider four types of privacy concerns, based on the affected stakeholder- the privacy of the infected individual, exposed individual, healthy or unexposed individual as well as the privacy of the businesses visited by the infected individual when they were contagious. 

For manual tracing, privacy of the infected individual as well as their contacts may be at stake if the interviews done by the contact tracers are recorded. The contact tracers enter the information collected into the database of the public health agency records and care should be taken to obtain the infected individual and their contacts’ consent before the identity and health information is shared with others. Typically, the names of the businesses are released on public health agency or government agency websites to alert possible contacts that the infected individuals did not know; for example, if the exposure to the virus occurred at a crowded bar, typically the name of the bar is posted on websites or released in the media - so the businesses do not have any privacy. 

There are two types of privacy risks to an individual when we consider exposure notification apps -  identity privacy (the individual will not want their identity to be disclosed without their consent) and location privacy (the individual would not want someone to be able to link the various locations they visited to figure out their location history, without their consent). By using privacy-preserving techniques, it is possible for smartphone apps to obfuscate actual location of users and provide both identity and location privacy. However, for exposure notification apps, it is necessary for the infected individual to disclose their location history so the app can automatically send a retrospective alert to possible contacts who may have been exposed to the infection due to their proximity to the infected individual when the infected individual was contagious; using privacy-preserving techniques, the location history could still be obfuscated to achieve some privacy, even for the infected individual. Since the location history of the infected individuals can be obfuscated, the identity of the businesses visited by the infected individuals when they were contagious are also not revealed, guaranteeing privacy to the businesses when infected individuals use privacy-preserving exposure notification apps. The exposed individuals do not need to reveal their location information until they get tested and are found to be infected also. However, the identity of the infected individual or the exposed individuals need not be revealed through the app, so they could achieve identity privacy through the app. The identity of the contacts could be revealed to the public health agency or their respective doctors, if the contacts reach out to them based on the exposure notification. Healthy and non-exposed individuals never receive any notifications from the app, need not contact a public health agency or their doctors and do not have to disclose their location history through the app. \cite{Rogue}

There are also security risks to the information collected by both manual tracers and the digital apps. Information collected by manual tracers as well as digital apps should be encrypted, when stored on a central server, so only authorized users have access. By using cryptographic techniques, it is possible to generate aggregate information from an encrypted database. All communications between smartphone apps and servers that are part of the digital contact advice system must also be encrypted. \cite{Patient} \cite{Samarajiva}

Additionally, manual contact tracers should also be trained in health privacy laws such as HIPAA, since they are collecting sensitive personal and health information from infected individuals as well as their contacts. Similarly, any software used for managing digital contact advice must be built using health privacy and exchange standards; it may be challenging for developers to be aware of all existing privacy and health information exchange laws since they may vary across states and countries.  

\definecolor{Green}{rgb}{0.58, 0.77, 0.49}

\begin{table}[!ht]
    \centering
    \begin{tabular}{|p{0.2\textwidth}|p{0.4\textwidth}|p{0.4\textwidth}|}
    \hline
    & Manual contact tracing & Digital contact advice\\
    \hline
    Privacy of infected individual 
    \cellcolor{Green}
    &
    None, especially if interviews are recorded
    &
    Good, if privacy techniques are used. Infected individuals need not reveal identity. Location history need not reveal identity of all locations they visited, thus providing some location privacy
    \\
    \hline
    Privacy of exposed individuals
    \cellcolor{Green}
    &
    None, especially if interviews are recorded
    &
    Good, if privacy techniques are used. Exposed individuals need not reveal identity or their location history
    \\
    \hline
    Privacy of healthy/non-exposed individuals
    \cellcolor{Green}
    &
    Good
    &
    Good
    \\
    \hline
    Privacy of businesses
    \cellcolor{Green}
    &
    None. Public health agencies may post names of businesses on websites or may release the information to the media.
    &
    Good, if privacy techniques are used so identity of locations visited by infected individuals need not be revealed
    \\
    \hline
    \end{tabular}
    \label{tab:privacy}
\end{table}
\section{Sociobehavioral Factors}

\textbf{Empathy and Trust}: Manual tracing involves human-to-human interactions, which provide a  rich connection. A well-trained tracer can use this personal connection to develop trust and to convince infected individuals to reveal details of their activities. There is no evidence that newly trained tracers are successful. Additionally, most manual contact tracers are local to the community and this knowledge might influence their social behavior even with those who are related to the exposed. Alternatively, exposure notification apps lack empathy, and rely on the willingness of the person who was notified to follow the recommendations provided by the app and reach out to public health agencies. 

\textbf{Risky Behavior (false negatives)}: If data collected by manual tracing is breached, sovereign immunity may protect public health agencies. It is also possible that contact tracers could be fired for bad behavior. Digital platforms can give rise to some of the risky behavior including spoofing data. In general, digital data platforms consist of three security risks - confidentiality, integrity, and availability. The confidentiality of the data becomes a privacy problem in the context of contact tracing data. Integrity involves spoofing the system by injecting different data samples maliciously or through other means. Availability of a digital contact tracing platform is relatively easy to compromise by attacks like DDOS.

\textbf{Misinformation}, \textbf{Risk of reporting errors (missing contacts; false positive exposures)}: Manual tracing offers  the opportunity to clarify misconceptions, since it is based on the notion of human conversation. It allows for different ways of misinformation suppression like inquiry, and manual verification, but does not completely prevent misinformation. Alternatively, with digital approaches, false information could be collected since multiple individuals may share the same phone. 

\textbf{Panic (false positives)}: False positives are highly unlikely in manual tracing as contacts are identified by the infected individuals, based on their recollection of their encounters. For digital approaches, however, given the potential for low risk proximity (eg, separation between individuals by a wall, transient interactions, etc), there is a higher probability of false positives. Even given the potential for false positives, public health authorities feel confident that digital approaches support, and don’t detract from manual contact tracing efforts.

\definecolor{Red}{rgb}{0.92, 0.82, 0.86}

\begin{table}[!ht]
    \centering
    \begin{tabular}{|p{0.2\textwidth}|p{0.4\textwidth}|p{0.4\textwidth}|}
    \hline
    & Manual contact tracing & Digital contact advice\\
    \hline
    Risky Behavior
    \cellcolor{Red}
    &
    If data is breached either by an insider or an outsider then there could be multiple concerns from a privacy and trust standpoint.
    &
    Digital platforms can give rise to some of the risky behavior including spoofing data which could harm some particular individuals/communities more than others.
    \\
    \hline
    Misinformation
    \cellcolor{Red}
    &
    MCT offers  the opportunity to clarify misconceptions. It is based on the notion of human conversation, hence it allows for different ways of misinformation suppression like inquiry, manual verification etc. However, it is still possible to propagate misinformation.
    &
    People often share one smartphone with other family members. They can leave their phone at charging stations hosting dozens of other devices.
    \\
    \hline
    Security of Information
    \cellcolor{Red}
    &
    It depends on the health authority, jurisdiction or country systems and infrastructure in place. Mostly variable depending upon the privacy laws enacted for that jurisdiction..
    &
    Similar to privacy in most of the context. If secure tools are used, this issue can be controlled to some extent. The risk that data collected for the purpose of contact tracing may be used for other purposes – or connected with other data sets to identify and potentially further profile individuals – is a central concern.
    \\
    \hline
    Socioeconomic Factors
    \cellcolor{Red}
    &
    At minimum need access to telephone so homeless and other indigent populations not included
    &
    Requires access to more advanced telephones (smartphones) or other connected devices, thus significant “at risk” populations may not be included (eg, those that live in highly dense public housing with poor resources, nursing homes, etc) 
    \\
    \hline
    Panic
    \cellcolor{Red}
    &
    Low risk, highly likely that exposures are “true exposures” since based on memory of the individual in question.
    Manual contact tracing occurs most often through human-to-human encounters, with the opportunity to clarify misconceptions, address worries, and express sympathy and other important affects.
    &
    High risk, given the potential for low risk proximity (eg, separation between individuals by a wall, transient interactions, etc)
    Technologies or apps may produce some false negatives or false positives, They need to be accurate enough that public health authorities feel confident that they support, and don’t detract from, contact tracing efforts.
    DCA intervention would affect all users regardless of circumstances (though some more than others) \cite{Kahn}
 \\
    \hline
    \end{tabular}
    \label{tab:sociobehavorial}
\end{table}
\section*{Discussion}
In an effective hybrid approach, digital tracing could provide a pre-populated list of potential contacts which can then be used by manual tracers during their interview with infected individuals. In the table that follows, we identified that using a hybrid solution could eventually strengthen privacy, Preserve data security, and confidentiality of contacts, if privacy preserving tools are considered as an aid to existing systems. It would further Support public health authorities and manual contact tracing efforts, as few repetitive processes can be automated  for faster triag. It would promote Interoperability with existing manual processes and across jurisdictions (local / state / national / international), as a mechanism to ensure the safe passage of goods and people, while  Eliminating the scope of human-error from memory recall more than a week ago. Contact tracing can scale bttr in a hybrid solution as the technology will be able to Proactively contain clusters before they grow exponentially, while also Reducing  the costs associated with scaling up extra manual contact tracers to meet the changes in demand.\cite{Dhillon}

\section*{The Decisions We Make Today} 
Action is urgently required. It has become clear that the coronavirus is not going away anytime soon. We need to see the virus before we can fight it, and right now we can’t see anything until it’s too late. In this case information isn’t just power, it’s survival. Contact tracing is one of the few ways we can prevent the disease on a massive scale. Contact tracing could allow us to return to some sense of normalcy in the coming months. As is becoming clear, that is an extremely valuable proposition, specifically if a hybrid solution is considered where digital interventions aid the current manual processes.Reconciling these imperatives will however, require thoughtful decisions about the types of technology we use, how it will be integrated in current systems,  and who will control the data generated. Focus on making transparency and consent the default settings is a necessary first step to getting it right. We have an unprecedented opportunity to implement a new and coherent data system that empowers individuals, health officials, and governments alike without creating surveillance state measures, while also keeping the empathy and human-centeredness at its core. There’s more to contact tracing than suppressing transmission. Contact tracing is also a great way to gather data, which can help scientists learn about the virus and be prepared for future pandemics. We need to make decisions around these issues fast. Our time to respond to the virus is running out. \cite{Pahwa}

\bibliographystyle{plain}
\bibliography{references}

\end{document}